\documentclass[reprint,showpacs,preprintnumbers,amsmath,amssymb,superscriptaddress,aps,prb]{revtex4-1}
\usepackage{graphicx}% Include figure files
\usepackage{epstopdf}
\usepackage{dcolumn}% Align table columns on decimal point
\usepackage{bm}% bold math

\newcommand{\ket}[1]{\ensuremath{\left|#1\right\rangle}}

\newcommand{\upstate}[0]{\ket{\uparrow} }
\newcommand{\downstate}[0]{\ket{\downarrow} }

\newcommand{\microev}[0]{\ensuremath{\mu}eV }

\begin{document}

\title{Detection and measurement of spin-dependent dynamics in random telegraph signals}

\begin{abstract}
A quantum point contact was used to observe single-electron fluctuations of a quantum dot in a GaAs heterostructure. The resulting random telegraph signals (RTS) contain statistical information about the electron spin state if the tunneling dynamics are spin-dependent. We develop a statistical method to extract information about spin-dependent dynamics from RTS and use it to demonstrate that these dynamics can be studied in the thermal energy regime. The tunneling rates of each spin state are independently measured in a finite external magnetic field. We confirm previous findings of a decrease in overall tunneling rates for the spin excited state compared to the ground state as an external magnetic field is increased.
\end{abstract}


\author{M. G. House}
\email{matthew.house@unsw.edu.au}
\altaffiliation{Present address: Centre for Quantum Computation and Communication Technology, University of New South Wales, Sydney, NSW 2052, Australia}
\affiliation{Department of Physics and Astronomy, University of California, Los Angeles, Los Angeles, California 90095, USA}
\author{Ming Xiao}
\affiliation{Key Laboratory of Quantum Information, Chinese Academy of Sciences, University of Science and Technology of China, Heifei 230026, People's Republic of China}
\author{GuoPing Guo}
\email{gpguo@ustc.edu.cn}
\affiliation{Key Laboratory of Quantum Information, Chinese Academy of Sciences, University of Science and Technology of China, Heifei 230026, People's Republic of China}
\author{HaiOu Li}
\affiliation{Key Laboratory of Quantum Information, Chinese Academy of Sciences, University of Science and Technology of China, Heifei 230026, People's Republic of China}
\author{Gang Cao}
\affiliation{Key Laboratory of Quantum Information, Chinese Academy of Sciences, University of Science and Technology of China, Heifei 230026, People's Republic of China}
\author{M. M. Rosenthal}
\affiliation{Department of Physics and Astronomy, University of California, Los Angeles, Los Angeles, California 90095, USA}
\author{HongWen Jiang}
\affiliation{Department of Physics and Astronomy, University of California, Los Angeles, Los Angeles, California 90095, USA}

\maketitle
\begin{figure}[tb]
\begin{center}\includegraphics[width=.95 \linewidth,keepaspectratio]{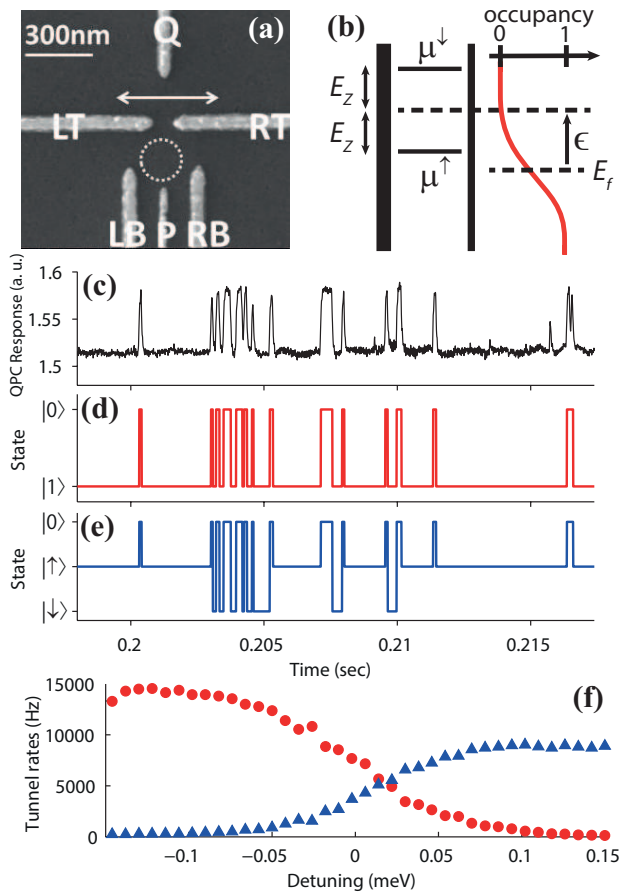}\end{center}
\caption{(a) SEM image of the active area of the device. A quantum dot is formed in the region highlighted by the dashed circle, connected by a tunnel barrier to a reservoir of free electrons at the lower right. A QPC channel through the upper half of the device is used to monitor the charge state of the quantum dot. (b) Schematic of the chemical potential levels of the quantum dot and occupancy of states in the lead. Indicated are the chemical potentials $\mu^\downarrow$ and $\mu^\uparrow$ for spin-up and spin-down electrons respectively, the Fermi level of the lead $E_f$, the detuning $\epsilon$, and the Zeeman energy $E_Z$. The red curve represents the Fermi distribution of occupied states in the lead. (c) Example of part of one RTS data trace. (d) Most-likely state sequence reconstructed from a two-state model fit to the RTS shown in (d). (e) Most-likely state sequence from a three-state model fit. (f) Overall tunneling rates $\Gamma_{IN}$ (red circles) and $\Gamma_{OUT}$ (blue triangles) as a function of detuning, determined by the two-state model at $B=0$ T.}
\label{fig:fig1composite}
\end{figure}

%\section{Introduction}
A random telegraph signal (RTS) is a natural phenomenon frequently observed in nanoscale solid state devices when an electron tunnels into and out of a trapping defect due to thermal fluctuations \cite{kandiah89, ghibaudo02}. An RTS can be produced in a controllable way by trapping an electron in an electrostatically defined quantum dot \cite{lu03}. The fluctuating electron influences the conductance of a nearby quantum point contact (QPC)\cite{buttiker90}, resulting in a conductance signal which switches back and forth between two distinct levels, low when an electron is present and high when it is not, as shown in Fig. \ref{fig:fig1composite} (c). RTS are also produced in experiments which drive an electron to a low-lying excited state, and serve as a signature of single-electron excitation \cite{xiao04, yin13}.

While an RTS exhibits only two distinct conductance levels, each corresponding to a charge state of the quantum dot, it is possible that more than two quantum states participate in the fluctuation. For example, the electron has a spin state which cannot be directly observed; nevertheless this extra ``hidden'' state may influence the statistics of the electron tunneling events in a measurable way. An existing strategy for finding statistical evidence for such states in RTS data has been to compute the full counting statistics (FCS) of the electron transits \cite{gustavsson09, flindt09}. Such analysis can reveal hidden structure in the RTS if multiple RTS are analyzed simultaneously \cite{gustavsson06}. The FCS approach does not take advantage of all of the information available in the signal and is subject to biases due to event reconstruction errors (see Supplementary Material). In a previous paper some of us proposed an analysis approach based on the hidden Markov model (HMM), which fits a rate equation model to the system to determine transition rates \cite{house09}. Here we present an adaptation of that approach based on Markov-Modulated Gaussian Process (MMGP) models \cite{roberts08}. These models account for noise in the signal so that it does not bias parameter estimates. They do not assign a definite state to the system at each point in time, only a probability of each state, which makes them robust against noise in the signal and the effect of the finite bandwidth of the measurement channel \cite{house09}. The models we use are very general and could be applied to study other types of quantum states with energy spacings similar to the thermal energy, such as valley states in Si quantum dots or hyperfine states in donors with large hyperfine interactions, \textit{e. g.} Bi donors in Si \cite{george10}.

The spin states of electrons in semiconductor quantum dots are a topic of current research interest in part because of their potential application for storing and manipulating information in classical information systems \cite{zutic04} and quantum information systems \cite{hanson07, loss98}. A challenge in studying electronic spins in quantum dots is that the spin cannot be observed directly but must be converted to an electrical signal for measurement \cite{elzerman04, hanson05, petta05b}. In this Letter we demonstrate that spin-dependent single-electron dynamics measurably influence the statistics of electron transition timings in RTS of a GaAs quantum dot. By applying statistical models to RTS we detect this influence and study spin-dependent dynamics at smaller Zeeman energies within the thermal regime. Our analysis reveals a spin dependence in the tunnel-out rates of electrons in GaAs quantum dots, similar to previous experimental findings on the tunnel-in rates \cite{amasha08}.

The experiment was performed on a quantum dot formed electrostatically in a GaAs/AlGaAs heterostructure, a device on which we have reported previously \cite{li12}. Fig. \ref{fig:fig1composite} (a) shows a scanning electron microscope (SEM) image of the surface gates. The quantum dot is defined in the area circled in the image by negative voltages applied to the five gates LT, RT, LB, RB, and P. Gates Q, LT, and RT form a QPC channel for sensing the charges on the quantum dot. The experiment was performed in a $^3$He refrigerator operating at a base temperature of 240 mK. The quantum dot was tuned so that the tunnel barrier between LB and LT was completely closed and the barrier between RB and RT was adjusted so that the tunneling rate between the quantum dot and the lead to the lower right was smaller than the bandwidth of the measurement channel (30 kHz). The capacitive coupling strength of gate P acting on the quantum dot was measured to be $\alpha = 0.022$ eV / V by a Coulomb diamond plot. To generate RTS data sets, $V_P$ was tuned so that the chemical potential of the dot was close to the Fermi level of the lead for the $N=0 \leftrightarrow N=1$ electron transition so that one electron tunneled on and off the dot, leaving it empty when the electron tunneled out. Each RTS trace was sampled at 131.1 kHz and collected for 7.6 seconds. After each RTS was collected, the voltage $V_P$ was stepped to change the de-tuning $\epsilon$ of the quantum dot's chemical potential relative to the Fermi level of the reservoir, and another RTS data set was taken. Each RTS was then analyzed independently by fitting statistical models to the data as described below.

A diagram of the chemical potential levels $\mu^\downarrow$ and $\mu^\uparrow$ of a single electron with two spin states are shown in Fig. \ref{fig:fig1composite} (b). The rates at which electrons tunnel in (out) of the dot are proportional to the fraction of occupied (unoccupied) states in the lead at the potential. When an external magnetic field $B$ is applied the potentials of two spin states are split by the Zeeman effect, resulting in different tunnel rates for the two spin states. The tunnel rates as a function of the detuning $\epsilon=\mu(B=0)-E_f$ are expected to obey
\begin{equation}
\label{eq:tunnelmodelin}
\Gamma^{\downarrow/\uparrow}_{IN}(\epsilon) = \Gamma^{\downarrow/\uparrow}_0 \exp[-\beta \epsilon] f(\epsilon \pm E_Z)
\end{equation}
\begin{equation}
\label{eq:tunnelmodelout}
\Gamma^{\downarrow/\uparrow}_{OUT}(\epsilon) = \Gamma^{\downarrow/\uparrow}_0 \exp[-\beta \epsilon] [1-f(\epsilon \pm E_Z)] + \Lambda_{OUT}
\end{equation}
where $\Gamma^{\downarrow/\uparrow}_0$ are the gross tunneling rates for the two spin states, $f$ is the Fermi distribution, $E_Z=g \mu_B B / 2$ the Zeeman energy, $\beta$ a factor accounting for the energy dependence of the tunneling rates \cite{maclean07}, and $\Lambda_{OUT}$ a term which accounts for a small back-action effect on which we have reported previously \cite{li12}. The $\pm$ in these equations is $+$ for spin-up and $-$ for spin-down. The $g$-factor in GaAs quantum wells is $g=-0.44$. By applying an in-plane magnetic field $|B|>0$ the energy levels of the two spin states are split and the two spin states have distinct tunneling rates near zero detuning.

\begin{figure*}[tb]
\begin{center}\includegraphics[width=7in,keepaspectratio]{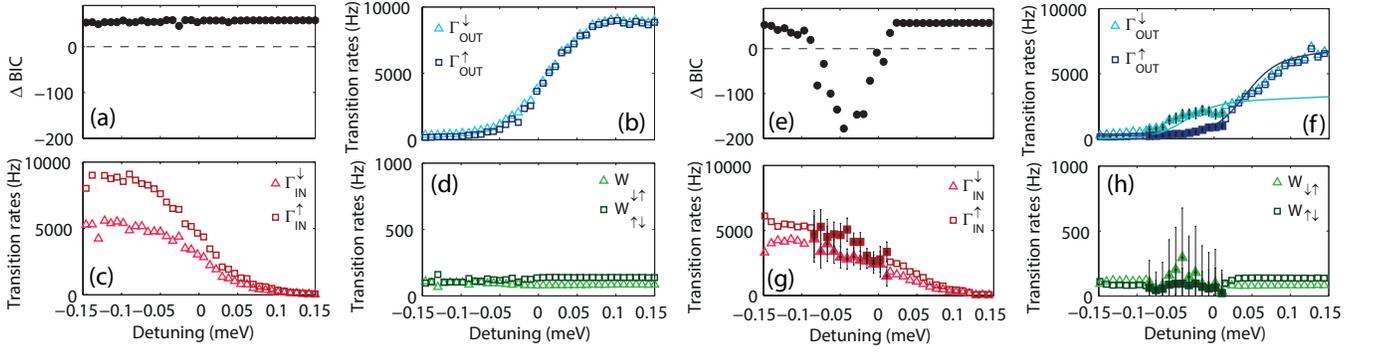}\end{center}
\caption{Results of fitting a three-state MMGP model to two sets of RTS. (a)-(d): magnetic field $B=0$ T. (e)-(h) $B=3$ T. (a) and (e) show the model selection statistic $\Delta BIC$ as a function of detuning $\epsilon$. Negative $\Delta BIC$ indicates that differences between two spins are statistically significant. (b) and (f) Tunnel-out rates $\Gamma^\downarrow_{OUT}$ and $\Gamma^\uparrow_{OUT}$ extracted from the three-state model fits. Those detuning points for which the three-state model is selected are emphasized by filled symbols and 90\% confidence intervals. (c) and (g) Tunnel-in rates $\Gamma^\downarrow_{IN}$ and $\Gamma^\uparrow_{IN}$. (d) and (h) Spin-flip transition rates $W_{\downarrow \uparrow}$ and $W_{\uparrow \downarrow}$. Solid lines in (f) indicate fits of Eq. (2) to those data points where the three-state model was selected.}
\label{fig:splitrates}
\end{figure*}

A MMGP is a statistical model with two parts: an observed signal (QPC current measurements) has statistical dependance on the state of an underlying system (the quantum dot). The state of the system is unknown and must be inferred from the signal. The state of quantum dot system is governed by the Markov equation
\begin{equation}
\label{eq:markov}
%\frac{d}{dt}\bm{p} = \bm{p} Q
d\bm{p}/dt = \bm{p} Q,
\end{equation}
where $\bm{p}$ is a row vector whose element $p_i(t)$ is the probability of the system being in state $i$ at time $t$, and $Q$ is a matrix whose element $Q_{ij}$ is the instantaneous probability of transition from state $i$ to state $j$. For each recorded data point the signal $I_{QPC}$ is taken from a Gaussian distribution whose mean $\mu_k$ depends on the state of the system at the time of the measurement (proportional to the number of electrons on the dot); the standard deviation $\sigma$ represents the experimental noise amplitude. For a given RTS the model parameters $Q$, $\mu_0, \mu_1,$  and $\sigma$ can be estimated by the Roberts-Ephraim algorithm \cite{roberts08}. The number of states the model system must be selected in advance.

The simplest rate equation model which can describe a RTS is a two-state model which has one state with one electron on the quantum dot, \ket{1} corresponding to mean signal level $\mu_1$, and one state with no electron, \ket{0} corresponding to $\mu_0$. In such a model the transition matrix has the form
\begin{equation}
\label{eq:twostateq}
Q=\left ( \begin{array}{cc} -\Gamma_{OUT} & \Gamma_{OUT} \\ \Gamma_{IN} & -\Gamma_{IN} \end{array} \right )
\end{equation}
where $\Gamma_{IN}$ and $\Gamma_{OUT}$ are respectively the mean tunneling rates for electrons in and out of the quantum dot. Fig. \ref{fig:fig1composite} (d) shows a state sequence reconstruction of the data in Fig. \ref{fig:fig1composite} (c) by fitting to the two-state model, as reconstructed by the Viterbi algorithm \cite{viterbi67}. The two-state model was fit to a sequence of RTS, each of which was taken at a different tuning of the quantum dot with respect to the Fermi level of the reservoir. The tunneling rates estimated by these model fits are shown in Fig. \ref{fig:fig1composite} (f). $\Gamma_{IN}$ is high, and $\Gamma_{OUT}$ low, at negative detunings because the occupancy of states in the lead is high; as the detuning is increased $\Gamma_{IN}$ decreases and $\Gamma_{OUT}$ increases with the decreasing occupancy of states in the lead. The maximum $\Gamma_{IN}$ is approximately twice the maximum of $\Gamma_{OUT}$ because there are two electron states for the system to tunnel into, $\Gamma_{IN} \approx \Gamma^\downarrow_{IN} + \Gamma^\uparrow_{IN}$, while the observed tunnel-out rate represents an average of the two spin states, $\Gamma_{OUT} \approx (\Gamma^\downarrow_{OUT} + \Gamma^\uparrow_{OUT})/2$.

The two-state model captures the aggregate behavior of the electron fluctuations but not any more complicated dynamics that may be present. To allow for spin-dependent effects we also construct a three-state model, which has one state with a spin-up electron \upstate, one state with a spin-down electron \downstate, and one state with zero electrons \ket{0}. The form of the transition matrix for this model is
\begin{equation}
\label{eq:threestateq}
Q=\left ( \begin{array}{ccc} -\Gamma^\downarrow_{OUT}-W_{\downarrow \uparrow} & W_{\downarrow \uparrow} & \Gamma^\downarrow_{OUT} \\
W_{\uparrow \downarrow} & -\Gamma^\uparrow_{OUT}-W_{\uparrow \downarrow} & \Gamma^\uparrow_{OUT} \\
\Gamma^\downarrow_{IN} & \Gamma^\uparrow_{OUT} & -\Gamma^\downarrow_{IN}-\Gamma^\uparrow_{OUT}  \end{array} \right )
\end{equation}
where \textit{e.g.} $\Gamma^\downarrow_{IN}$ is the rate for an electron to tunnel into the dot in the state \downstate, $W_{\downarrow \uparrow}$ is the rate for a spin-flip transition from \downstate to \upstate, and so on. States \downstate and \upstate both correspond to the same mean QPC signal level $\mu_1$ while \ket{0} corresponds to level $\mu_0$. By fitting this model to an RTS we can extract estimates for the spin-dependent transition rates $\Gamma^\downarrow_{OUT}$, $\Gamma^\uparrow_{OUT}$,$\Gamma^\downarrow_{IN}$,$\Gamma^\uparrow_{IN}$, $W_{\downarrow \uparrow}$, and $W_{\uparrow \downarrow}$. Shown in Fig. \ref{fig:fig1composite} (e) is the most-likely sequence of states in the three-state model fit to the data in Fig. \ref{fig:fig1composite} (c). The model preferentially assigns the state \upstate to longer dwells of the electron on the dot and \downstate to shorter ones.

To determine whether the spin-dependent dynamics included in the three-state model are justified by the data or not, we compare the goodness-of-fit of the two models to each RTS by their Bayesian Information Criterion (BIC) statistic \cite{kass95}. The BIC of each model is computed as $BIC=-2\log(\hat{\mathcal{L}})+K \log(N)$, where $\mathcal{\hat{L}}$ is the maximum-likelihood value of the model (evaluated by the Roberts-Ephraim algorithm), $K$ is the number of degrees of freedom in the model, and $N$ is the number of data points. For a given data set the preferred model is the one with the lower value of $BIC$. Under this selection criterion the two-state model (having fewer degrees of freedom) will be selected unless there is significant evidence in the data for the three-state model, as reflected in its higher likelihood. If the three-state model is selected we may confidently say that the RTS data is not adequately explained as a two-state system and that there must be a third state with unique dynamics present.

The results of three-state model fits to a series of RTS with applied magnetic field $B=0$ T are shown in Fig. \ref{fig:splitrates} (a)-(d). Fig. 2 (a) shows the difference in BIC between the two-state model and the three-state model. At every detuning point the two-state model is selected ($\Delta BIC$ is positive), indicating that there are no statistically significant spin-dependent effects. This is expected, since with no applied magnetic field the two spin states are degenerate and should have the same tunneling rates. Fig. \ref{fig:splitrates} (e)-(h) show the same results taken with applied magnetic field $B=3$ T. In this case, at moderate negative detunings the selection criterion favors the three-state model ($\Delta BIC$ is negative). It is in this region where the two tunnel-out rates $\Gamma^\downarrow_{OUT}$ and $\Gamma^\uparrow_{OUT}$ are different and there are a large enough number of electron transitions observed to make the difference statistically significant. At positive detunings most of the electron transitions involve the spin-up state and  there are not enough instances of the spin-down state to make its rates distinguishable. At large negative detunings the tunnel-out rates become small and there are relatively few transitions to either state. Because of the nature of the way the tunnel rates influence the RTS statistics, the individual tunnel-out rates can be measured with greater accuracy than the tunnel-in rates (see Supplementary Materials). In this experiment the spin-flip transition rates $W_{\downarrow\uparrow}$ and $W_{\uparrow\downarrow}$ were too small to measure, but an upper bound can be placed on them as shown by the confidence intervals in Fig. \ref{fig:splitrates} (h). All of the uncertainties are statistical and could be reduced by taking longer RTS traces.

\begin{figure}[tb]
\begin{center}\includegraphics[width=\linewidth,keepaspectratio]{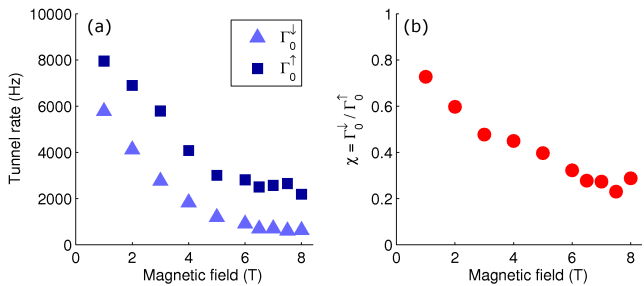}\end{center}
\caption{(a) Gross tunnel-out rates for the two spin states, $\Gamma_0^\downarrow$ and $\Gamma_0^\uparrow$, as a function of applied magnetic field. These were determined by fits to the rates extracted from a series of RTS such as shown in Fig. \ref{fig:splitrates} (f) for $B=3$ T. (b) Ratio of the gross tunnel-out rates for the two spin states, $\Gamma_0^\downarrow/\Gamma_0^\uparrow$, as a function of applied magnetic field. The tunneling rate for spin-down electrons decreases relative to the spin-up tunneling rate as the Zeeman energy difference between the two states increases.}
\label{fig:rates_vs_b}
\end{figure}

The solid lines in Fig. \ref{fig:splitrates} (f) represent fits of Eq. \ref{eq:tunnelmodelout} to the tunnel rates for each spin, using only those data points where the three-state model was selected. The same analysis was applied to similar data sets taken with applied magnetic fields from $B=1$ T to $B=8$ T. At $B=1$ T the Zeeman energy splitting ($2 E_Z = 25 \microev$) is comparable to the thermal energy ($k_B T = 21 \microev$), and the spin states are distinguishable in this regime (see Supplementary Materials). The gross tunneling rates $\Gamma^\downarrow_0$ and $\Gamma^\uparrow_0$ extracted from these fits are plotted in Fig. \ref{fig:rates_vs_b} (a) as a function of magnetic field, which shows that for $|B|>0$ the overall tunneling rate for spin-down electrons is smaller than that for spin-up, even after accounting for the Fermi distribution of occupied states in the lead and the energy dependence of the tunneling rate. This difference in tunneling rates increases as the magnetic field is increased and the ratio of the tunneling rates of the two spins, $\Gamma_0^{\downarrow} / \Gamma_0^\uparrow$, plotted in Fig. \ref{fig:rates_vs_b} (b), is seen to decrease monotonically with increasing $B$ as reported by Amasha, \textit{et al}. \cite{amasha08}. The spin selective tunneling effect may be due to an enhancement of the $g$-factor of the electron in the lead to which the electron is coupled, but it is still not well understood \cite{stano10}. Here we are able to see the effect primarily in the tunnel-out rates instead of tunnel-in rates, and at lower energies than previously reported.

In order to verify that there are no additional states participating in the electron fluctuations we used two additional models: a model containing one one-electron state and two zero-electron states (an unphysical model according to our interpretation), and a model containing two states for each electron number (four states total). In every case, the BIC for these models were greater than for the models described previously, indicating that there is not significant evidence for state configurations other than those we addressed.

In summary, we have described a technique for analysis of RTS and detecting structure the underlying system such as a difference in tunneling rate for two different states of a quantum dot with the same electron number. We used this technique on data obtained from a GaAs quantum dot and detected an extra ``hidden'' state at negative detunings when a magnetic field is applied. We identified the extra state in this system with the spin of the electron because its behavior is consistent with Zeeman physics in the presence of an applied magnetic field. The approach is very general and could be used to study other effects in RTS that cannot be observed directly, such as other spin configurations, valley states, or hyperfine states. Because RTS occur at the thermal energy scale, this type of analysis can reveal such effects happening at smaller energies than other experimental methods. The same type of modeling could also be applied to detect hidden structure in quantum jumps, which have the same statistical nature as RTS \cite{sauter86,vijay11}. In the future the classical MMGP model can be extended to apply to fully quantum mechanical processes, which will allow an experimenter to estimate quantum coherent effects in stochastic processes \cite{housedissertation}.

The authors thank M. Y. Simmons for valuable discussions about this manuscript. This work was supported by the U.S. Army Research Office (w911NF-11-1-0028). MGH acknowledges support from the UCLA Graduate Division Dissertation Year Fellowship. The authors from USTC acknowledge support from the National Fundamental Research Program (Grant No. 2011CBA00200)， National Natural Science Foundation, and Chinese Academy of Sciences．

\bibliographystyle{apsrev4-1} % Choose Phys. Rev. style for bibliography
\bibliography{hmmustcpaper}

\end{document}